\documentclass[12pt,epsf,a4paper]{article}
\usepackage[english]{babel}
\usepackage{latexsym,epsfig}
\usepackage{amsmath}

0

\begin{document}
\baselineskip 6,5mm

\def\II{\relax{\rm 1\kern-.35em1}}
\def\IP{\relax{\rm I\kern-.18em P}}
\renewcommand{\theequation}{\thesection.\arabic{equation}}
\csname @addtoreset\endcsname{equation}{section}
\def\muh{\hat \mu}
\def\nuh{\hat \nu}
\def\rhoh{\hat \rho}
\def\lambdah{\hat \lambda}
\def\sigmah{\hat \sigma}
\def\tauh{\hat \tau}

\begin{flushright}
hep-th/0011204 \\
IC/2000/174
\end{flushright}

\begin{center}

{}~\vfill

{ {\LARGE {Glueball Masses for the Deformed Conifold Theory}}}

\vspace{20 mm}

{\bf {\large {Elena C\'aceres}}} {\large and} {\bf{\large{Rafael Hern\'{a}ndez}}}$^{\dag}$

\vspace{8 mm}

{\em The Abdus Salam International Center for Theoretical Physics \\
Strada Costiera, 11. $34014$ Trieste, Italy}
\vspace{16 mm}

\end{center}


\begin{center}
{\bf Abstract}
\end{center}
  
\vspace{2 mm}

We obtain the spectrum of glueball masses for the ${\cal N}=1$ non-conformal 
cascade theory whose supergravity dual was recently constructed by Klebanov 
and Strassler. The glueball masses are calculated by solving the 
supergravity equations of motion for the dilaton and the two-form 
in the deformed conifold background.

\vspace{42 mm}

{\footnotesize \dag}\hspace{1 mm}{\footnotesize{\ttfamily e-mail address: caceres@ictp.trieste.it, 
rafa@ictp.trieste.it}}

\newpage


\section{Introduction}

The original AdS/CFT correspondence \cite{ads/cft}-\cite{Witten1} was generalized in 
\cite{Kachru}-\cite{Morrison} for branes at conical singularities (see also 
\cite{Gubser}-\cite{Klebanov:2000me} for related work). If instead 
of locating D3-branes in a flat transversal space they are placed at the vertex of a six 
dimensional cone with base a five dimensional Einstein manifold $X_5$, one is lead 
to conjecture that type IIB string theory on $AdS_5 \times X_5$ is dual to the low 
energy limit of the worldvolume theory on the D3-branes at the singularity. In 
particular, a set of $N$ D3-branes at the conifold singularity results on a 
${\cal N} =1$ superconformal field theory with  $SU(N) \times SU(N)$ gauge group \cite{KW1}. 
Conformal invariance can be broken 
using  fractional D3-branes which are allowed to appear in certain  
singular spaces \cite{Gimon}-\cite{KN}. Fractional D3-branes arise from D5-branes wrapped at 
the collapsed two-cycles of the singularity. In particular, the addition of $M$ fractional 
branes at the singular point of the conifold modifies the gauge group of the field theory to 
$SU(N+M) \times SU(N)$. 
This theory was first investigated in \cite{KN} to leading order in $M/N$. This solution was 
then completed to all orders in \cite{KT}. However, the result found by the authors 
of \cite{KT} contains an infrared singularity. In \cite{KS}, Klebanov and Strassler constructed 
a non singular solution valid from the ultraviolet to the infrared. They showed 
that the theory undergoes a series of Seiberg dualitites as 
the gauge group is sucesively broken. Far in the infrared 
the D3-branes disappear,
and the theory is ${\mathcal N}=1$ $SU(M)$ with no matter; this theory 
exhibits confinement, domain walls and screening. Using the supergravity description 
of gauge theories, as in \cite{Witten-g}-\cite{gb2}, we will 
in the present letter find glueball masses
 for ${\cal N}=1$ $SU(M)$ Yang-Mills. In section 2 we will 
review some of the aspects of the solution constructed in 
\cite{KS}, emphasizing those relevant to our calculation. In section 3 we analyze the 
equations of motion of type IIB supergravity, and their final form after dimensional 
reduction on a sphere. The equations for the type IIB dilaton and the two-form fields 
are numerically solved in section 3 to determine the mass spectra of the corresponding  
glueball modes. Finally, in section 4, we present some conclusions and comment on 
future perspectives\footnote{After completion of this paper, reference \cite{krasnitz} 
appeared, which partially overlaps with the present work.}.


\section {Review of Klebanov-Strassler's Solution}{\label{review}} 

The solution recently found by Klebanov and Strassler arises from the 
study of D3-branes at a singular space. Originally, 
the authors of \cite{KW1} studied the conformal field theory on D3-branes at a Calabi-Yau 
singularity dual to type IIB on a $AdS_5 \times T^{11}$ background. The corresponding 
gauge theory is 
${\mathcal N} =1$ supersymmetric with  $SU(N) \times SU(N)$ 
gauge group, and matter content $A_{i}$,$B_{i}$ where $i=1,2$. The chiral fields 
transform as $({\bf N}, 
{\bar {\bf N}})$ and  $({\bar {\bf N}},{\bf N})$ respectively. 
The superpotential is $W = \lambda \epsilon^{ij} \epsilon_{kl} \hbox {Tr } A_i B_k A_j B_l$ . 
In the presence of M fractional branes the superpotential and matter
content are the same but the gauge group changes to $SU(N+N)\times SU(N)$. The chiral
superfields are  now in the representation $({\bf N+M}, {\bar {\bf N}})$. The supergravity 
equations corresponding to this situation were solved, to leading 
order in $M/N$, in \cite{KN}, where the relative gauge coupling $g_1^{-2}-g_2^{-2}$ was 
shown to run logarithmically. This approximation was completed to all orders in \cite{KT}. 
In this solution  a logarithmic harmonic function warps the conifold,
\begin{equation}
ds^2 = \frac {r^2}{L^2 \sqrt{\ln (r/r_s)}} dx_n dx_n + \frac {L^2 \sqrt{ \ln (r/r_s)}}{r^2} 
dr^2 + L^2 \sqrt{ \ln (r/r_s)} ds^2_{T^{11}}.
\label{1}
\end{equation}
  
These fractional D3-branes at the singularity are D5-branes wrapped over the 
collapsed $S^2$ of $T^{11}$. D5-branes are sources for 
the magnetic R-R three-form flux through the $S^3$ cycle of $T^{11}$ and thus 
adding M fractional D3-branes implies that  the supergravity dual 
of this field theory will involve $M$ units of three-form flux,
\begin{equation}
\int_{S^3} F_3 = M,
\label{2}
\end{equation}
in addition to  $N$ units of five-form flux coming from the D3-branes,
\begin{equation}
\int_{T^{11}} F_5 = N.
\label{3}
\end{equation}
This non vanishing three-form is  responsible of the conformal symmetry 
breaking. As a consequence of this flux, the two-form $B_2$ is no longer constant; it 
develops a radial dependence \cite{KN},
\begin{equation}
\int_{S^2} B_2 \sim M e^{\phi} \ln (r/r_s),
\label{4}
\end{equation}
while the dilaton remains constant.
  
The supergravity solution obtained in \cite{KN} was completed in \cite{KT} taking into 
account the back reaction of $H_3 = dB_2$ and $F_3$ on other fields. This exact solution 
exhibits an effect which was hidden to leading order in $M$: since 
$F_5 = dC_4 +B_2 \wedge F_3$, $F_5$ also acquires a radial dependence,
\begin{equation}
F_5 = {\cal F}_5 + \star {\cal F}_5,
\label{5}
\end{equation}
where 
\begin{equation}
{\cal F}_5 = {\cal K}(r)  \hbox{vol }(T^{11}) = 
(N + a g_s M^2 \ln (r/r_0)) \hbox{vol }(T^{11}),
\label{6}
\end{equation}
with $a$ a constant of order unity. But there is a novelty arising from this 
solution: the five-form flux present at the  
ultraviolet scale $r=r_0$ may disappear once we reach the scale $r=\tilde{r}$, where 
${\cal K}(\tilde{r})=0$. This phenomenon can be related to the fact that the flux 
$\int_{S^2} B_2$ is not a periodic variable in the supergravity solution, because as 
this flux goes through a period, ${\cal K}(r) \rightarrow {\cal K}(r) -M$, so that 
the five-form flux is decreased by $M$ units. This decrease represents a renormalization 
group cascade, that was identified in \cite{KS} as a form of Seiberg duality \cite{Seiberg}.
  
However, the metric (\ref{1}), representing the logarithmic renormalization group cascade, contains 
a naked singularity at $r=r_s$, which is the point where the harmonic function vanishes, $h(r_s)=0$. 
From a physical point of view the singularity represents the end point of the cascade as 
the theory flows to the infrared, because negative values of $N$ are unphysical. And it is this 
singularity in the metric (\ref{1}), that  demands a 
modification, at least in the 
infrared, of the solution. This non singular correction was found in \cite{KS}, where 
it was argued that the conifold should be replaced by a deformed conifold, 
\begin{equation}
\sum_{i=1}^4 z_i^2 = - 2 \: \hbox{det}_{ij} \, z_{ij} = \epsilon^2,
\label{7}
\end{equation}
with the singularity removed through the blow up of the $S^3$ in $T^{11}$. A simple argument 
in favor of this suggestion comes from the origin of this singularity; it 
can be related to the divergent energy of the $F_3$ field. As there are $M$ units of 
flux of $F_3$ through the $S^3$ in $T^{11}$, when $S^3$ shrinks to zero size $F_3$ 
diverges. However, if $S^3$ is kept of finite size, as in the deformed conifold, there is 
no need for $F_3$ to diverge.
  
A deeper argument comes from a detailed field theory analysis, which shows that the 
spacetime geometry should be modified by the strong dynamics of the infrared limit of 
the field theory. The authors of \cite{KS} showed how the $U(1)$ (${\bf Z}_{2M}$, to be 
more precise) R-symmetry is broken to a ${\bf Z}_2$ symmetry. This is indeed the symmetry 
left unbroken on the deformed conifold (\ref{7}), $z_k \rightarrow - z_k$, instead of the 
original $U(1)$ $z_k \rightarrow e^{i \alpha} z_k$. The metric was then shown to be of 
the form 
\begin{equation}
ds_{10}^2 = h^{-1/2} (\tau) dx_n dx_n + h^{1/2} (\tau) ds_6^2,
\label{8}
\end{equation}
with $ds_6^2$ the metric of the deformed conifold. In the basis $\{\tau, g^{i=1,\ldots,5} 
(\psi,\theta_1,\theta_2,\phi_1,\phi_2) \}$ of reference  
\cite{Minasian} this metric becomes diagonal,
\begin{equation}
\begin{split}
ds_6^2 &= \frac {1}{2} \epsilon^{4/3} K(\tau) \Big[ \frac {1}{3 K^3(\tau)} [d \tau^2 
+ (g^5)^2] + \cosh^2 \left( \frac {\tau}{2} \right) [(g^3)^2 + (g^4)^2]  \\
&+ \sinh^2 \left( \frac {\tau}{2} \right) [(g^1)^2+(g^2)^2]  \Big],
\label{9}
\end{split}
\end{equation}
where
\begin{equation}
K(\tau) = \frac {(\sinh(2 \tau)-2 \tau)^{1/3}}{2^{1/3} \sinh(\tau)}.
\label{10}
\end{equation}
    
The harmonic function in (\ref{8}) is given by the integral expression 
\begin{equation}
h(\tau) = \alpha \frac {2^{2/3}}{4} \int_{\tau}^{\infty} dx \frac {x \coth x -1}{\sinh^2 x}
(\sinh(2x)-2x)^{1/3},
\label{13}
\end{equation}
which  cannot be evaluated in terms of elementary or special functions. The constant 
$\alpha$ is $\alpha \sim (g_s M)^2$.
  
The solution contains a five-form and three-form flux.  $F_5 = {\cal F}_5 + 
\star {\cal F}_5$ is  given by;
\begin{equation}
{\cal F}_5=  g_s M^2 l(\tau) g^1 \wedge g^2 \wedge g^3 \wedge g^4 \wedge g^5, 
\label{f5}
\end{equation}
and   
\begin{equation}
\star {\cal F}_5 = g_s M^2 \frac {2 \, l(\tau)}{15 \, K^2(\tau) h^2(\tau) \sinh^2(\tau)
\epsilon^{8/3}} dx^0 \wedge dx^1 \wedge dx^2 \wedge dx^3 \wedge d \tau,
\label{15}
\end{equation}
while the three-form is
\begin{eqnarray}
G_3= & &F_3 +\frac{i}{g_s}H_3\\
   = & & M \big\{ g^5\wedge g^3 \wedge g^4 + d[ F(\tau) (g^1 \wedge g^3 + g^2 \wedge g^4)]
\nonumber \\
   & & + i d[f(\tau) g^1 \wedge g^2 + k(\tau) g^3 \wedge g^4] \big\}, \nonumber \\
\end{eqnarray}
with the functions
\begin{eqnarray}
l(\tau)& = & \frac {\tau \coth \tau -1}{4 \sinh^2 \tau} (\sinh 2 \tau - 2 \tau), \nonumber\\
f(\tau)& = & \frac {\tau \coth \tau -1}{2 \sinh \tau}   (\cosh \tau - 1), \nonumber\\
k(\tau)& = & \frac {\tau \coth \tau -1}{2 \sinh \tau}   (\cosh \tau + 1), \nonumber\\
F(\tau)& = & \frac {\sinh \tau - \tau}{2 \sinh \tau}. \nonumber\\
\label{16}
\end{eqnarray} 
  
In the far infrared, through the chain of Seiberg dualities that drop the size of the 
gauge group factors by $M$ units, the D3-brane  goes to  zero and  
only the $M$ fractional D3-branes remain. Thus, at the bottom of the cascade we are left 
with a pure ${\cal N}=1$ Yang-Mills theory 
with $M$ isolated vacua. In this letter we will use the above 
supergravity description of this theory, proposed by Klebanov and Strassler, to study the 
glueball mass spectra of this theory by solving numerically the equations of motion describing 
the propagation of supergravity fields along the worldvolume of the branes.


\section{Supergravity Equations}

In this section we will write down the dimensional reduction of the linearized equations of 
motion of type IIB supergravity in the background described in section \ref{review}.
The bosonic sector of the type IIB supergravity multiplet contains a graviton 
$g_{\muh \nuh}$, a dilaton $\Phi$, a zero-form R-R field C, 
two tensors --the NS-NS and R-R fields $B_{\muh \nuh}$ and 
$C_{\muh \nuh}$-- and a R-R four-form $C_{\sigmah \lambdah \tauh \nuh}$. In the present
notation carets denote indices that run over ten dimensions, greek indices 
denote four dimensional space and latin indices run over the internal space.
Expanding the type IIB equations of motion \cite{Schwarz} 
in background (dotted fields) and fluctuations we get,
\begin{eqnarray}
D^{\muh}\partial_{\muh} \Phi& =& \frac{\kappa ^2}{24} 
{\dot G}_{\rhoh \sigmah \tauh} G^{\rhoh \sigmah \tauh},  \\
D^{\muh}G_{\muh \rhoh \sigmah}& =& - \frac {2 i \kappa}{3} [F_{\rhoh \sigmah \lambdah \tauh \nuh} 
{\dot G}^{ \lambdah \tauh \nuh} + {\dot F}_{\rhoh \sigmah \lambdah \tauh \nuh} G^{
\lambdah \tauh \nuh}] + \partial^{\muh} \Phi {\dot G}_{\muh \rhoh \sigmah}, \\
R_{\muh \nuh}&=& \frac {\kappa^2}{6} F_{\muh \rhoh \sigmah \lambdah \tauh } 
{\dot F}_{\nu}^{\: \: \rhoh \sigmah \lambdah \tauh } + 
\frac{\kappa^2}{8}[ \hbox {Re}({\dot G}_{\muh}^{\: \:\rhoh \sigmah}G_{\nuh \rhoh \sigmah}) 
- \frac{1}{6}g_{\muh \nuh}{\dot G}^{ \sigmah \lambdah \tauh}{\dot G}_{\sigmah \lambdah \tauh} 
\nonumber \\
&-&\frac{1}{6} {\dot g}_{\muh \nuh} {\dot G}^{ \sigmah \lambdah \tauh} G_{\sigmah 
\lambdah \tauh}], \\
F_{\muh_1 \muh_2 \muh_3 \muh_4 \muh_5 } & =& \frac {1}{5 \, !}
\varepsilon_{\muh_1 \muh_2 \muh_3 \muh_4 \muh_5 ...\muh_{10}}F^{\muh_6....\muh_{10}}
\end{eqnarray}

The presence of a non zero three-form in the  background  
implies the coupling of  the dilaton, metric and two form. 
In general this should require solving the system of coupled equations. However, 
these equations can be simplified by expanding the excitations in 
spherical harmonics \cite{hs}. Formally, one should expand in harmonics over the exact 
background. Nevertheless, since we are interested in calculating glueball masses, {\it i.e.}, 
in confining effects, and these occur near the bottom of the cascade where the  
the $S^2$ shrinks to zero but the $S^3$
does not we can  expand in Kaluza-Klein 
modes over the $S^3$. The expansion for the dilaton is, 
\begin{equation}
\Phi(x,y)= \sum \phi(x)^k Y(y)^k, 
\label{dex}
\end{equation}
while for the two-form
\begin{eqnarray}
A_{\mu \nu} (x,y) & = & \sum a_{\mu \nu}^{I_1} (x) Y^{I_1}(y), \nonumber \\
A_{\mu m} (x,y)   & = & \sum [ a_{\mu}^{I_5} (x) Y_m^{I_5} (y) + a_{\mu}^{I_1} (x) 
D_m Y^{I_1} (y) ], \nonumber \\
A_{mn} (x,y)      & = & \sum [ a^{I_{10}} (x) Y_{[mn]}^{I_{10}} (y) + a^{I_5} (x) 
D_{[m} Y_{n]}^{I_5} (y) ],
\label{aex} 
\end{eqnarray} 
where $x= x^\mu$ are four dimensional worldvolume coordinates are
$y=x^i$ are coordinates over the three sphere. The Lorentz type gauges 
$D^{m} A_{mn}=0$, $D^{m} A_{m \mu} =0$ can be used in order to fix 
$a_{\mu}^{I_1}=a^{I_5}=0$ in (\ref{aex}). 
  
First, we will consider the equation (3.1) for the dilaton which  only requires scalar spherical 
harmonics. Substituting the expansion 
(\ref{dex}) into the field equation for the dilaton we see that 
the $s$-wave of the kinetic side becomes $D^{\mu} \partial_{\mu} \phi(x) \, Y(y)$, 
where we have omitted the $k=0$ index. However, the three-form of the Klebanov-Strassler 
solution \cite{KS} lives only on the internal coordinates so its spherical harmonic 
expansion involves $Y_{[m n]}$ but no scalar harmonics. Thus, 
the $s$-wave equation for the dilaton becomes  
$D^{\mu} \partial_{\mu} \phi (x)=0$ or, explicitly,
\begin{equation}
\frac {1}{\sqrt{g}} \partial_{\mu} [ \sqrt{g} \partial_{\nu} \phi g^{\mu \nu} ] = 0. 
\label{dileq}
\end{equation}
  
In order to study the propagation of the supergravity two-form along the worldvolume of the 
branes we should consider the $D^{\hat{\alpha}} G_{\hat{\alpha} \mu \nu}$ part of equation 
(3.2). Using
\begin{equation}
C_{\mu m n p} = \sum \phi_{\mu}^{I_5} (x) \varepsilon_{mnp}^{\: \: \: \: \: \: \: \: qr} 
D_q Y_r^{I_5} (y)
\end{equation}
together with  (\ref{dex}) and (\ref{aex})  the $s$-wave equation of motion 
for the two-form in the worldvolume  is
\begin{equation}
\frac {3}{\sqrt{g}} g_{\mu \mu'} g_{\nu \nu'} \partial_{\alpha} [ \sqrt{g} 
\partial_{[\alpha'} a_{\mu'' \nu'']} g^{\alpha \alpha'} g^{\mu'' \mu'} g^{\nu'' \nu'} ] 
= - \frac {2i \kappa}{3} {\dot F}_{\mu \nu \rho \sigma \tau} \partial^{[\rho} a^{\sigma \tau ]},
\label{aa}
\end{equation}
where $[ \: \: ]$ denotes antisymmetrization with strength one, and $a_{\mu \nu}$ is the 
complexified two-form.


\section{Glueball Mass Spectra}
  
In this section we will study the discrete spectrum of glueball masses
arising from propagation in (\ref{8}) of the type IIB dilaton field, and the complex antisymmetric 
field $a_{\mu \nu}$. The $SU(N+M) \times SU(N)$ conifold theory has a $U(1)_R$ global 
symmetry \cite{KT}. In the present case, the deformation of the conifold breaks the 
$U(1)_R$ to ${\bf Z}_2$. Therefore, one expects a massless glueball in this theory. However, 
this massless glueball will couple to some combination of supergravity fields\footnote{We 
thank I. Klebanov for comments on this point.}, and thus we do not expect to find it in our 
analysis of the glueball spectra.
  
In \cite{GKP,Witten1} the correspondence between the $AdS_5 
\times S^5$ background and primary chiral fields correlators was made explicit. 
This correspondence should, in principle, be modified for the deformed conifold background 
since the space is not asymptotically $AdS$ anymore. However, as argued in \cite{KS}, 
only operators with $\Delta < \frac {3}{2} M$ can propagate all the way to 
$\tau =0$. For this type of operators one expects that the correspondence should 
not be greatly modified. Thus, one expects that the dilaton will couple to the 
dimension four operator $\hbox {Tr } 
F_{\mu \nu} F^{\mu \nu}$, so that the spin zero glueball masses, which can be derived 
from the two point function of this operator, will arise from propagation of the dilaton. 
Similarly, the supergravity two-form should couple to a dimension six  
operator, {\it i.e.}, to the spin one glueball.

In order to obtain a pure glue gauge theory we should take the 
$g_s M \rightarrow 0$ limit. In this limit the curvature of the space is
 large and  we are outside of the region where supergravity has small corrections. 
This situation is not new when dealing with string duals of pure glue theories. Thus, 
in the spirit of \cite{gb1,gb2} we proceed with the calculation and expect the 
corrections to  masses to be small. With this caveat let us now consider the 
equation for the dilaton. Expanding the dilaton field in plane waves, 
$\phi (\tau,x) = f(\tau) e^{ik.x}$, so that 
a mode of momentum $k$ has a mass  $m^2 = - k^2$ and using the metric (\ref{8})
the dilaton equation (\ref{dileq}) becomes   
\begin{equation}
3.2^{1/3} \frac {d}{d \tau} \Big[ (\sinh (2 \tau) - 2 \tau)^{2/3} 
\frac {df}{d \tau} \Big] - (k^2 \epsilon^{4/3}) \sinh^2 (\tau) h(\tau) f = 0.
\label{b}
\end{equation}
    
As a boundary condition we will 
require that near the origin the function $f$ must be smooth, so that $df/d \tau =0$ at 
$\tau = 0$. The asymptotics of $f(\tau)$ as $\tau \rightarrow \infty$ is obtained by 
demanding normalizability of the states. Since for large $\tau$,
\begin{equation}
\sqrt{g} = 2^{-5} 3^{-1} 
\epsilon^4 \sinh^2(\tau) h^{1/2}(\tau) \sim \tau e^{2/3 \tau}
\end{equation}
convergence of the integral 
$\int \! \sqrt{g} \: \: |\Phi|^2$ signals an exponential behaviour of the solution 
near infinity, $f \sim e^{n \tau}$, as can be suspected from direct inspection 
of (\ref{b}). Changing  variables to $f(\tau) = \psi(\tau) e^{n \tau}$, the wave 
equation (\ref{b}) becomes
\begin{equation}
\begin{split}
&6[\sinh(2 \tau) -2 \tau]^{2/3} \, \psi'' + [6n 2^{1/3} (\sinh (2 \tau) - 2 \tau)^{2/3} \\
&+ 2^{7/3} (\cosh ( 2 \tau)-1)(\sinh(2 \tau) - 2 \tau)^{-1/3}] \, \psi' 
+ [3 n^2 2^{1/3} (\sinh (2 \tau)-2 \tau)^{2/3} \\
&+ 2^{7/3} n (\cosh(2 \tau) -1)
(\sinh (2 \tau) -2 \tau)^{-1/3} - (k^2 \epsilon^{4/3}) \sinh^2(\tau) h(\tau)] \, \psi = 0,
\label{c}
\end{split}
\end{equation}
This equation  can be solved for large values of $\tau$ by $\psi(\tau) = c_1 e^{-n \tau} + 
c_2 e^{- \frac {1}{3} (3n +4) \tau}$ which implies  
$f(\tau) = c_1 + c_2 e^{-4/3 \tau}$. Normalizability requires $c_1=0 $ and we 
can fix $c_2=1$. Thus, at infinity $f \sim e^{-4/3 \tau}$. 
Equation (\ref{b}) becomes then an eigenvalue problem; we want to find the values of $k$ 
for which the equation has solutions matching the 
boundary conditions at $0$ and $\infty$. We have done this numerically using 
a shooting technique \cite{gb1}: boundary conditions at infinity are taken 
as initial conditions and $k$ is adjusted  so that when numerically integrating 
(\ref{b}) $f$ is also smooth at $\tau=0$. 
We should note that, numerically, infinity is taken as some large value of 
$\tau$ such that $\tau \gg k^2$. On the other hand, since the boundary 
condition 
requires that $f$ goes exponentially fast to zero at infinity, there is a 
numerical bound that does not allow us to push the numerical infinity as far 
as we want. This explains why we are restricted to obtain only the 
first two eigenvalues. 
The glueball masses 
thus obtained are shown in Table~1. They are measured in units of $\epsilon^{4/3}$ which, 
as explained in \cite{KS}, sets the four dimensional mass scale of the field theory.

\begin{table}[hbt]
\centering

\begin{tabular}{|l|c|}     \hline\hline
   {\bf State} &  {\bf (Mass)$^2$} \\ \hline
   $0^{++}$    &  9.78 \\
   $0^{++*}$   &  33.17 \\
    \hline\hline
\end{tabular}

\caption{Mass (squared) of the spin zero glueball and its first 
excited state obtained from supergravity.}

\label{tab1}
\end{table}
  
It is interesting to note that by varying the origin of integration one
can see confining effects occurring in a small region close to 
the bottom of the cascade. Past this small region the spectrum 
becomes continuous signaling a conformal behaviour. 
  
Next we will study the equation of motion for the complexified two-form $a_{\mu \nu}$. 
Using again the metric 
(\ref{8}) and the five-form flux through the worldvolume, (\ref{15}), 
equation (\ref{aa}) becomes 
\begin{equation}
\begin{split}
3.2^{1/3}  
\frac {d}{d \tau} &\Big[ h(\tau) (\sinh(2 \tau) -2 \tau)^{2/3} \frac {d f_{\mu \nu}} 
{d \tau} \Big] - (k^2 \epsilon^{4/3}) \sinh^2 (\tau) h^2(\tau) f_{\mu \nu} \\
&= - \frac {8 i \kappa}{15} g_s M^2 l(\tau) \epsilon^{-2} \Big[ \varepsilon_{\mu \nu \tau \alpha 
\beta} \epsilon^{-2/3} \frac {d f_{\alpha \beta}}{d \tau} + 
i k \frac {\epsilon^{2/3} h(\tau) }{6 K^2(\tau)}
\varepsilon_{\mu \nu x \alpha \beta} f_{\alpha \beta} \Big],
\label{bb}
\end{split}
\end{equation}
where we have impossed solutions to be of the form $a_{\mu \nu} (\tau,x) = 
f_{\mu \nu}(\tau) e^{ik.x}$, and summation is implied over the $\alpha$ and 
$\beta$ indices. We should note that as a consequence of the non constant 
five-form background of the solution of Klebanov and Strassler, the second order 
differential operator (\ref{aa}) can not be factorized into two first order 
operators, as in \cite{hs}. Thus, we should take care explicitly of the derivatives 
in the right hand side of (\ref{aa}). This requires the complex decomposition 
$f_{\mu \nu} (\tau) = b_{\mu \nu} (\tau) + i c_{\mu \nu} (\tau)$. Impossing the gauge fixing 
condition $a_{\mu \tau} =0$, so that the two-form has no components along the 
radial direction, equation (\ref{bb}) becomes 
a system of coupled equations. Similarly as in the dilaton case, the glueball masses 
can be obtained by numerically solving this system. The values are shown in Table~2.

\begin{table}[hbt]
\centering

\begin{tabular}{|l|c|}     \hline\hline
   {\bf State} &  {\bf (Mass)$^2$} \\ \hline
   $1^{--}$    &  14.05 \\
   $1^{--*}$     &  42.90 \\
    \hline\hline
\end{tabular}

\caption{Mass (squared) of the spin one glueball and its first 
excited state obtained from supergravity.}

\label{tab2}
\end{table}


\section{Discussion}

In this paper we have computed the dilaton and two-form excitations 
in the deformed conifold background recently constructed by Klebanov and 
Strassler \cite{KS}. Via the correspondence between supergravity and field theory, this 
corresponds to determining masses for glueballs in the dual effective field theory.
Unlike the $AdS_5 \times S^5$ case, the Klebanov-Strassler background
contains non trivial three-form and five-form fluxes. 
Far in the infrared the three-form flux prevents the three-cycle of the base 
from collapsing. We obtained the linearized type IIB equations by doing 
a Kaluza-Klein decomposition on this $S^3$. We 
solved the resulting eigenvalue problem by numerically integrating these 
equations. This method is exact and allows us to find the first two excited 
states for the dilaton and two-form fluctuations. A WKB approximation could also be 
used to solve the equations. This  approach was used in \cite{1}-\cite{3} to 
determine glueball masses for several finite temperature supergravity models. A comparison of WKB
results with those of \cite{gb1}, where a numerical method was used, shows that agreement increases 
for excited states. This is to be expected since the WKB 
approximation improves for large values of the masses.
This might be at the root of the small discrepancy, when comparing ratios of eigenvalues, between 
our results for the spin zero glueball and those in \cite{krasnitz}, 
where the WKB approximation was used. Due to the fact that the equations 
involve extremely divergent functions our numerical method is 
not able to find higher excited states and we cannot compare with 
\cite{krasnitz} for states where the WKB approximation is more reliable.
It would be interesting to compare with lattice results 
but we are not aware of any lattice computation for supersymmetric glueballs.
  
The present computations can be generalized to higher spin 
glueballs. It would also be interesting to study mixed supergravity 
states; one would then expect to find a massless glueball as a consequence of the 
breaking of the $U(1)_R$ symmetry.


\vspace{8 mm}

{\bf Acknowledgements}

It is a pleasure to thank Stathis Tompaidis for useful discussions. We also 
thank Pilar Hern\'andez and Igor Klebanov for correspondence. This research is 
partly supported by the EC contract no. ERBFMRX-CT96-0090.

\newpage


\end{document}